\documentstyle[12pt]{article}
\begin{document}

\thispagestyle{empty}
{}~ \hfill TECHNION-PH-96-1 \\
\vspace{1cm}

\begin{center}
{\large \bf  Two-body Cabibbo-suppressed  Decays of Charmed Baryons}\\
{\large \bf  into Vector Mesons and into Photons}\\
\vspace{1.50cm}
{\bf Paul Singer and Da-Xin Zhang}\\
\vspace{0.50cm}
Department of Physics, Technion -- Israel Institute 
of Technology, Haifa 32000,
Israel\\
\vspace{1.5cm}
{\bf Abstract}
\end{center}

\vspace{1cm}
\noindent 
The heavy quark effective theory and the factorization approximation
are used to treat the Cabibbo-suppressed decays
of charmed baryons to vector mesons, 
$\Lambda_C\rightarrow  p{\rho^0}, p\omega$, 
$\Xi_C^{+,0}\rightarrow\Sigma^{+,0}\phi, 
\Sigma^{+,0}{\rho^0}, \Sigma^{+,0}\omega$ and  $\Xi_C^{0}\rightarrow\Lambda\phi,
\Lambda\rho, \Lambda\omega$. 
The input from two recent experimental results on $\Lambda_C$ decays
allows the estimation of the branching ratios
for these modes, which turn out to be between $10^{-4}$ and $10^{-3}$.
The long distance contribution of these transitions
via  vector meson dominance to the radiative weak processes
$\Lambda_C\rightarrow p\gamma$, $\Xi_C\rightarrow\Sigma\gamma$ 
and $\Xi_C^0\rightarrow\Lambda\gamma$
leads to quite small branching ratios, $10^{-6}-10^{-9}$;
the larger value holds if a sum rule between the coupling constants
of the vector mesons is broken. 

\vspace{1.2cm}
\noindent 
{\rm PACS number(s):}  13.30.Eg, 12.40.Vv, 14.20.Kp

\newpage
The study of the charmed baryon decays has intensified
during the last  few years on both the experimental and theoretical
levels\cite{review}.
In particular, the recent measurements with the CLEO-II detector
at the Cornell Electron Storage Ring(CESR) of the 
formfactor ratio in semileptonic
decay $\Lambda_C\rightarrow\Lambda e^+\nu$\cite{cleo1} 
and of the branching ratio of $\Lambda_C\rightarrow p\phi$\cite{cleo2},
provide information whose usefulness
transcends the specific processes studied in these experiments.
In the present work, we focus
on a group of Cabibbo-suppressed two-body
nonleptonic decays of the charmed baryons $\Lambda_C$ and $\Xi_C$ 
into a light baryon plus a light-flavor neutral vector
meson, which were not treated previously, making use of 
information provided by the two experiments of Ref. \cite{cleo1} 
and \cite{cleo2}. 

The Cabibbo-suppressed decays of the charmed hadrons
are described at the quark level  by the effective Hamiltonian
\begin{equation}
{\cal H}_{eff} = \sum_{q=d, s}
\displaystyle\frac{G_F}{\sqrt{2}}
V_{uq}^*V_{cq} a_2 \bar{q}\gamma^\mu(1-\gamma_5) q
\bar{u}\gamma_\mu(1-\gamma_5) c,
\label{heff}
\end{equation}
where $a_2$ is a  combination of  Wilson coefficents, 
$a_2=c_1(\mu=m_c)+c_2(\mu=m_c)/N_c$\cite{bsw}. 
In (\ref{heff}) only
the leading contribution in the large $N_C$ limit
is retained.
In order to  evaluate  decay amplitudes induced by the
effective Hamiltonian ${\cal H}_{eff}$ at the  hadronic level
we shall use  the factorization approximation\cite{bsw}
and we concentrate now on those decays in which $(q\bar{q})$ 
materializes to a vector meson. For instance,
the decay amplitude of $\Lambda_C\rightarrow p\phi$ is  given by
\begin{equation}
{\cal M}(\Lambda_C\rightarrow p\phi)=
\displaystyle\frac{G_F}{\sqrt{2}} V_{us}^*V_{cs} a_2 
<\phi|\bar{s}\gamma^\mu (1-\gamma_5) s|0>
<p|\bar{u}\gamma_\mu (1-\gamma_5) c|\Lambda_C>.
\label{pphi}
\end{equation}
While the matrix element for the vector meson is related to
its decay constant defined by
\begin{equation}
<\phi|\bar{s}\gamma^\mu s|0>=i f_\phi m_\phi\epsilon^{*\mu}_\phi,
\label{three}
\end{equation}
the baryonic matrix element can be parameterized by six
form factors $f_i$ and $g_i$ (i=1,2,3)\cite{cheng1}:
\begin{equation}
\begin{array}{rcl}
<p|\bar{u}\gamma_\mu (1-\gamma_5)  c|\Lambda_C>&=&\bar{u}_p(P_2)
[(f_1(q^2)\gamma_\mu-i\displaystyle\frac{f_2(q^2)}{m_{\Lambda_C}}\sigma_{\mu\nu}q^\nu
+\displaystyle\frac{f_3(q^2)}{m_{\Lambda_C}}q_\mu)\\[4mm]
&&\hspace{-2cm}
-(g_1(q^2)\gamma_\mu-i\displaystyle\frac{g_2(q^2)}{m_{\Lambda_C}}\sigma_{\mu\nu}q^\nu
+\displaystyle\frac{g_3(q^2)}{m_{\Lambda_C}}q_\mu)\gamma_5
]u_{\Lambda_C}(P_1)
\end{array}
\label{hme1}
\end{equation}
with $q_{\mu}=(P_1-P_2)_{\mu}$.

We use  the heavy quark effective theory, 
which is considered to be especially suitable
for the $\Lambda_C$ decays\cite{hqet,kk}.
Treating the $c$ as the heavy-quark, the matrix element 
in (\ref{hme1}) is expanded in  $1/m_C$ and in the following we
keep only the first term of the expansion.
Then only two independent
formfactors survive and (\ref{hme1}) is cast into the form
\begin{equation}
<p|\bar{u}\gamma_\mu (1-\gamma_5)  c|\Lambda_C>=
\bar{u}_p(P_2)
[F_1(q^2)+F_2(q^2)
\displaystyle\frac{{\not\! P_1}}{m_{\Lambda_C}}]\gamma_\mu (1-\gamma_5)
u_{\Lambda_C}(P_1).
\label{hme2}
\end{equation}
The formfactors in (\ref{hme1})  and in (\ref{hme2}) are then related by
\begin{equation}
\begin{array}{llll}
&f_1(q^2)=F_1(q^2)+\displaystyle\frac{m_p}{ m_{\Lambda_C}}F_2(q^2), 
& f_2(q^2)=-F_2(q^2) ,& f_3(q^2)= F_2(q^2),\\[4mm]
&g_1(q^2)=F_1(q^2)+\displaystyle\frac{m_p}{ m_{\Lambda_C}}F_2(q^2), 
& g_2(q^2)=-F_2(q^2),& g_3(q^2)= F_2(q^2).\\[4mm]
\label{relat}
\end{array}
\end{equation}
The  relations in (\ref{relat}) are expected to hold most likely near
the zero-recoil point $q^2_m\equiv (m_{\Lambda_C}-m_p)^2\simeq 1.8{\rm GeV}^2$
of the semileptonic decays. 
Since we need here the form factors in the region of $\sim 1 {\rm GeV}^2$,
we shall assume it is appropriate to use a pole behavior
for the  extrapolation :
\begin{equation}
\begin{array}{rcl}
f_i(q^2)&=&f_i(q_m^2)\displaystyle\frac{1-q_m^2/M_{D^*}^2}{1-q^2/M_{D^*}^2},
\\[5mm]
g_i(q^2)&=&g_i(q_m^2)
\displaystyle\frac{1-q_m^2/M_{D_1}^2}{1-q^2/M_{D_1}^2},\\[5mm]
\end{array}
\label{qbehav}
\end{equation}
where $M_{D^*}=2.007$GeV and $M_{D_1}=2.423$GeV are the 
masses of the 
lowest lying mesons which interpolate the vector and the axial
vector currents, respectively.

Now we turn to the  experimental information\cite{cleo1}
on   the form factor ratio 
in $\Lambda_C\rightarrow \Lambda e^+\nu$. Although the absolute
value of $F_1$,  $F_2$ defined in (\ref{hme2})
is not  measured yet, their ratio is determined to be
\begin{equation}
R\equiv \displaystyle\frac{F_2}{F_1}=-0.25\pm 0.16
\label{r}
\end{equation}
in the semileptonic decay.
We remark that in view of the limited statistics, it was found in 
\cite{cleo1} that
this result  is not sensitive to the $q^2$ behavior
assumed for the formfactors.

Now we assume that as consequence of  the SU(3)-flavor 
symmetry for the light quarks
the ratio (\ref{r})  holds also for the matrix elements of 
 the decays  $\Lambda_C\rightarrow p$ and $\Xi_C\rightarrow\Sigma,\Lambda$. 
Hence, we are now in the position to derive the
decay amplitude  (2) in the approximation of (5) and we arrive at
\begin{equation}
{\cal M}(\Lambda_C\rightarrow p\phi)=
\displaystyle\frac{G_F}{\sqrt{2}} V_{us}^*V_{cs} a_2 
i f_\phi m_\phi\epsilon^{*\mu}_\phi
\bar{u}_p(P_2)
[\gamma_\mu(a-b\gamma_5)+2(x-y\gamma_5)P_{1\mu}]
u_{\Lambda_C}(P_1),
\label{pphi2}
\end{equation}
where
\begin{equation}
\begin{array}{rcl}
a&=&f_1(m_\phi^2)+\displaystyle
\frac{m_p+m_{\Lambda_C}}{m_{\Lambda_C}}f_2(m_\phi^2),\\[4mm]
b&=&g_1(m_\phi^2)+\displaystyle
\frac{m_p-m_{\Lambda_C}}{m_{\Lambda_C}}g_2(m_\phi^2),\\[4mm]
x&=&-\displaystyle\frac{1}{m_{\Lambda_C}}f_2(m_\phi^2),\\[4mm]
y&=&-\displaystyle\frac{1}{m_{\Lambda_C}}g_2(m_\phi^2).\\[4mm]
\end{array}
\end{equation}
Similiar formulas  hold when the  $\phi$  meson
is replaced by a ${\rho^0}$ or an $\omega$ meson in the final state,
where an additional factor $1/\sqrt{2}$ arises due to the
quark content of the meson  ${\rho^0}$ or  $\omega$ .
In the process of $\Xi_C^0\rightarrow \Lambda\phi ({\rm or} ~{\rho^0}, \omega)$ 
there exists another factor $\sqrt{1/6}$ to account for
the difference of  the flavor-spin suppression
 for the light quarks\cite{sing}.

We proceed now to calculate the decays listed in Table I.
Firstly, we use the experimental data of CLEO\cite{cleo2}
which measures the branching ratio 
Br($\Lambda_C\rightarrow p\phi)=(1.06\pm 0.33)\times10^{-3}$
as an input, thus determining the unknown product
$|a_2F_1(m_{\phi}^2)|$.
This permits  to calculate with the model
the transitions 
$\Lambda_C\rightarrow  p{\rho^0}, p\omega$, 
$\Xi_C^{+,0}\rightarrow\Sigma^{+,0}\phi, 
\Sigma^{+,0}{\rho^0}, \Sigma^{+,0}\omega$ and  $\Xi_C^{0}\rightarrow\Lambda\phi,
\Lambda\rho, \Lambda\omega$. 
Since the experimental uncertainty  in (\ref{r}) is
large, we present in Table I the  values for 
$R=-0.09$, $R=-0.25$, $R=-0.41$. It turns out that
the dependence on the ratio $R$ in the considered range
is weak, the variation in the calculated branching ratios
 being less than $10\%$.
The uncertainty in the results of Table I is then due solely
to the precision achieved in the determination of 
Br($\Lambda_C\rightarrow p\phi)$\cite{cleo2}.
The partial decay widths  are calculated
using  the helicity representation of the amplitudes\cite{helicity}
and thus we included in Table I also the predictions of the
model for transverse/longitudinal ratios in these decays, which
are found to vary about $20\%$ in the range considered for $R$.
In the calculations we used for  $f_V(V=\rho,\omega~ {\rm or } ~\phi)$ the
values determined from the leptonic decays of the vector mesons\cite{pdg},
 $f_\rho^2=0.047{\rm GeV}^2$, 
$f_\omega^2=0.038{\rm GeV}^2$, $f_\phi^2=0.055{\rm GeV}^2$.

As an alternative, we could have refrained from using $\Lambda_C\rightarrow p\phi$
as input , attempting to calculate it as well, by using
$a_2=-0.55\pm 0.1$ from the overall fit\cite{bsw} to nonleptonic $D$ and $D_s$
decays, and assuming a ``reasonable" value for $f_1(q_m^2)$.
With the above value for $a_2$ and taking, for example, $f_1(q_m^2)=0.9\pm 0.1$,
we find Br($\Lambda_C\rightarrow p\phi)=(0.74\pm 0.32) \times 10^{-3}$ 
(for $R=-0.25$). 
This is in good
agreement with the measured value\cite{cleo2} and gives 
strong support to the approach presented here.

In view of the appropriateness of the model discussed
here for decays of the charmed baryons to vector mesons,
we consider also its application to radiative weak decays 
in conjunction with vector meson dominance(VMD).
Since it is clear now\cite{burdman} that
the short distance contribution from $c\rightarrow u\gamma$
to the radiative decays of  charmed particles is negligible,
it is important to devise reliable models for the
long distance one.
We adopt here the model which has been employed recently by
Deshpande {\it et al.}\cite{bvmd2} and by Eilam {\it et al.}\cite{svmd}
to estimate the long distance ``t-channel" contribution of the vector
mesons to the radiative transitions $b\rightarrow s\gamma$ and 
$s\rightarrow d\gamma$(see also Ref. [14]).
Using now for the charm sector the derivation steps outlined in \cite{bvmd2,svmd},
the appropriate part of the effective Hamiltonian
for the radiative weak decays is
\begin{equation}
{\cal H}_{eff}^{VMD}
=-\displaystyle\frac{eG_F}{\sqrt{2}}a_2\displaystyle\frac{1}{m_C}
[V_{us}^*V_{cs}(-\displaystyle\frac{1}{3}f_\phi^2)
+V_{ud}^*V_{cd}(-\displaystyle\frac{1}{2}f_{\rho^0}^2+
\displaystyle\frac{1}{6}f_\omega^2)]
\epsilon_{\gamma}^{*\mu}q^\nu
\bar{u}\sigma_{\mu\nu}(1+\gamma_5) c.
\label{hvmd}
\end{equation}
To treat the newly
encountered  matrix element of  $\sigma_{\mu\nu}(1+\gamma_5) $,
we use the heavy quark effective scheme with the c-quark
only treated as heavy. Then, similiarly to Cheng {\it et al.}[15], we obtain
\begin{equation}
<p|\bar{u}\sigma_{\mu\nu}(1+\gamma_5)  c|\Lambda_C>=
\bar{u}_p(P_2)
[F_1(q^2)+F_2(q^2)\displaystyle\frac{{\not\! P_1}}{m_{\Lambda_C}}]\sigma_{\mu\nu}
(1+\gamma_5)  
u_{\Lambda_C}(P_1).
\label{hme3}
\end{equation}

Next, we use again the measured result of Br$(\Lambda_C\rightarrow p\phi)$
\cite{cleo2}
and of  $R$\cite{cleo1},
in order to calculate the VMD contributions\cite{bvmd2,svmd,vmd}
to the  processes  
$\Lambda_C\rightarrow p\gamma$, $\Xi_C^{+,0}\rightarrow\Sigma^{+,0}\gamma$ and 
$\Xi_C^0\rightarrow\Lambda\gamma$.
We are still faced, however,  with the question
of the $q^2$ dependence of the $f_V(q^2)$ couplings.
It is customary to assume that the $q^2$
variation of the $f_\rho$, $f_\omega$ is small and one may
take for the vector-meson-photon couplings at  $q^2=0$,
 $f_\rho(m_{\rho}^2)\simeq f_\rho(0)$, $f_\omega(m_{\omega}^2)\simeq f_\omega(0)$.
We may reasonably  assume also  $f_\phi(m_{\phi}^2)\simeq f_\phi(0)$.
Since in Eq (11) we use $V_{ud}^*V_{cd}\simeq -V_{us}^*V_{cs}$,
the amplitude for the processes considered is
proportional to the quantity $C'_{VMD}\equiv
-\displaystyle\frac{1}{3}f_\phi^2+\displaystyle\frac{1}{2}f_{\rho^0}^2-
\displaystyle\frac{1}{6}f_\omega^2$.
It has been pointed out already in Ref \cite{svmd}
that by using the values determined in the leptonic decays
for $f_V(0)$ a near cancellation occurs in $C'_{VMD}$.
This cancellation is due to the combination of $SU(3)$-flavor
symmetry with the GIM relation and occurs at a level below $10\%$.
As a result, all the considered radiative decays are
reduced by more than two orders of the magnitude.
Denoting the suppressed branching ratios obtained from
the combined contribution of $\phi$, $\omega$, $\rho$ by
the subscript $SR$(sum rule), we find:
\begin{equation}
\begin{array}{rcl}
Br(\Lambda_C\rightarrow p\gamma)_{SR}&=&
1.8\times 10^{-9}, 2.3\times 10^{-9}, 3.1\times 10^{-9},\\
Br(\Xi_C^+\rightarrow \Sigma^+\gamma)_{SR}&=&
3.5\times 10^{-9}, 4.5 \times 10^{-9}, 6.2\times 10^{-9},\\
Br(\Xi_C^0\rightarrow \Sigma^0\gamma)_{SR}&=&
1.0\times 10^{-9}, 1.3\times 10^{-9}, 1.8\times 10^{-9},\\
Br(\Xi_C^0\rightarrow \Lambda\gamma)_{SR}&=&
0.16\times 10^{-9}, 0.21\times 10^{-9}, 0.28\times 10^{-9}
\end{array}
\end{equation}
for $R=-0.09$, $-0.25$ or $-0.41$.
On the other hand, the possiblility exists that
a certain variation occurs in $f_V(q^2)$ between $q^2=0$
and $q^2=m_V^2$. For instance, there is strong
evidence that $f_\psi(q^2)$ varies considerably
between $q^2=m_\psi$ and $q^2=0$,$f_\psi^2$ being
reduced in this range by a factor of $6$\cite{bvmd2,svmd}.
Hence one should also consider the 
possiblility that the above mentioned 
cancellation is avoided. Since there is no
accurate model for the $f_V(q^2)$ variation,
we take as alternative the rates resulting if only the
 $\rho\leftrightarrow\gamma$ is considered. 
This leads to
\begin{equation}
\begin{array}{rcl}
Br(\Lambda_C\rightarrow p\gamma)_{\rho}&=&
0.73\times 10^{-6}, 0.93\times 10^{-6}, 1.3\times 10^{-6},\\
Br(\Xi_C^+\rightarrow \Sigma^+\gamma)_{\rho}&=&
1.4\times 10^{-6}, 1.7\times 10^{-6}, 2.5\times 10^{-6}\\
Br(\Xi_C^0\rightarrow \Sigma^0\gamma)_{\rho}&=&
0.41\times 10^{-6}, 0.53\times 10^{-6}, 0.73\times 10^{-6}\\
Br(\Xi_C^0\rightarrow \Lambda\gamma)_{\rho}&=&
0.65\times 10^{-7}, 0.97\times 10^{-7}, 1.1\times 10^{-7}
\end{array}
\end{equation}
for $R=-0.09$, $-0.25$ or $-0.41$.
These figures may be compared with a previous
calculation\cite{constitute} using bremsstrhlung from
W-exchange diagrams on the quark level.
Branching ratios of the order of $10^{-5}$ were
obtained for these processes, however, as the authors pointed
out, large uncertainties are involved.

In summary,
we have presented the first calculation of the Cabibbo-suppressed processes 
$\Lambda_C\rightarrow  p{\rho^0}, p\omega$, 
$\Xi_C^{+,0}\rightarrow\Sigma^{+,0}\phi, 
\Sigma^{+,0}{\rho^0}, \Sigma^{+,0}\omega$ and  $\Xi_C^{0}\rightarrow\Lambda\phi,
\Lambda\rho, \Lambda\omega$
by using the factorization approximation,  heavy
quark effective theory and experimental input from 
recent experiments.  We found branching ratios
between $10^{-4}$ and $10^{-3}$ for these
modes, which brings their detection into the realm of feasibility in the near future.
The model accounts correctly for the observed $\Lambda_C\rightarrow p\phi$.
Then, using vector meson dominance for the long distance
contribution, we calculated the radiative processes
$\Lambda_C\rightarrow p\gamma$, $\Xi_C^{+,0}\rightarrow \Sigma^{+,0}\gamma$,
and $\Xi_C\rightarrow \Lambda\gamma$ in the same model.
If a certain GIM-type sum rule holds for the vector-meson-photon
couplings these transitions are strongly suppressed
to the level $10^{-8}-10^{-9}$. Otherwise,
individual vector mesons contribute branching ratios
of the order of $10^{-6}$.
Detection or limits on these modes would thus test the validity
of interesting theoretical models.

This research  is supported in part by Grant 5421-3-96
from the Ministry of
Science and the Arts of Israel. The work
of   P.S. has also been supported in part by the Fund for Promotion of
Research at the Technion.
We thank Professor G. Eilam for helpful remarks.

\newpage
\pagestyle{empty}
\vskip -0.2cm
\noindent
{\bf Table I.} 
Predictions on branching ratios and the ratio of polarized
 \\[-2mm]
decay rates $\Gamma_L/\Gamma_T$.\\
\begin{tabular}{cc||c|c|c} 
\hline\hline
Process&&$R=-0.09$&$R=-0.025$&$R=-0.41$\\ 
\hline\hline
$\Lambda_C\rightarrow p\phi$&BR&$10^{-3}$(input)
&$10^{-3}$(input) &$10^{-3}$(input) \\ [-2mm]
& $\Gamma_L/\Gamma_T$
&1.18 &1.29 &1.42\\ \hline
$\Lambda_C\rightarrow p{\rho^0}$&BR
&$0.42\times 10^{-3} $&$0.44\times 10^{-3}$&
$0.47\times 10^{-3}$\\ [-2mm]
& $\Gamma_L/\Gamma_T$
&2.24 &2.50 &2.76\\ \hline
$\Lambda_C\rightarrow p\omega$&BR
&$0.34\times 10^{-3} $&$0.36\times 10^{-3}$&
$0.38\times 10^{-3}$\\ [-2mm]
& $\Gamma_L/\Gamma_T$
&2.17 &2.42 &2.67\\ 
\hline
\hline
$\Xi_C^+\rightarrow \Sigma^+\phi$&BR& 
$1.8\times 10^{-3}$
&$1.8 \times 10^{-3}$& $1.7\times 10^{-3}$\\[-2mm] 
& $\Gamma_L/\Gamma_T$
&1.04 & 1.14&1.25\\
\hline
$\Xi_C^+\rightarrow \Sigma^+{\rho^0}$&BR
&$0.80 \times 10^{-3}$
&$0.82 \times 10^{-3}$&$0.83 \times 10^{-3}$\\ [-2mm]
& $\Gamma_L/\Gamma_T$
&2.06 & 2.29&2.55\\
\hline
$\Xi_C^+\rightarrow \Sigma^+\omega$&BR
&$0.65 \times 10^{-3}$
&$0.66 \times 10^{-3}$&$0.67 \times 10^{-3}$\\ [-2mm]
& $\Gamma_L/\Gamma_T$
&1.98 & 2.21&2.46\\
\hline
\hline
$\Xi_C^0\rightarrow \Sigma^0\phi$&BR& 
$0.52\times 10^{-3}$
&$0.50 \times 10^{-3}$& $0.48\times 10^{-3}$\\ [-2mm]
& $\Gamma_L/\Gamma_T$
&1.04 & 1.14&1.25\\
\hline
$\Xi_C^0\rightarrow \Sigma^0{\rho^0}$&BR
&$0.23 \times 10^{-3}$
&$0.23 \times 10^{-3}$&$0.24 \times 10^{-3}$\\ [-2mm]
& $\Gamma_L/\Gamma_T$
&2.06 & 2.29&2.55\\
\hline
$\Xi_C^0\rightarrow \Sigma^0\omega$&BR
&$0.19 \times 10^{-3}$
&$0.19 \times 10^{-3}$&$0.19 \times 10^{-3}$\\[-2mm] 
& $\Gamma_L/\Gamma_T$
&1.98 &2.21&2.46\\
\hline
\hline
$\Xi_C^0\rightarrow\Lambda\phi$&BR& 
$0.93\times 10^{-4}$
&$0.92 \times 10^{-4}$& $0.90\times 10^{-4}$\\ [-2mm]
& $\Gamma_L/\Gamma_T$
&1.19 & 1.31&1.44\\
\hline
$\Xi_C^0\rightarrow\Lambda{\rho^0}$&BR
&$0.39 \times 10^{-4}$
&$0.40 \times 10^{-4}$&$0.41 \times 10^{-4}$\\ [-2mm]
& $\Gamma_L/\Gamma_T$
&2.28 & 2.54&2.83\\
\hline
$\Xi_C^0\rightarrow\Lambda\omega$&BR
&$0.31 \times 10^{-4}$
&$0.32 \times 10^{-4}$&$0.34 \times 10^{-4}$\\ [-2mm]
& $\Gamma_L/\Gamma_T$
&2.20 & 2.46&2.74\\
\hline\hline
\end{tabular}


\newpage


\begin{thebibliography}{99}
\bibitem{review}
J. G. K\"orner and H. W. Seibert,
Ann. Rev. Nucl. Part. Sci. 41, 511 (1991).

\bibitem{cleo1}
G. Crawford {\it et al.} (CLEO Collaboration), Phys. Rev. Lett.  {75}, 624 (1995).

\bibitem{cleo2}
J.P. Alexander {\it et al.} (CLEO Collaboration),  report CLNS 95/1343.

\bibitem{bsw}
M. Baur, B. Stech and M. Wirbel, Z. Phys. {C34}, 103 (1987).

\bibitem{cheng1}
H.-Y. Cheng and B. Tseng, Phys. Rev.  D46, 1042 (1992);\\
P.Zenczykowski, Phys. Rev.  D50, 402 (1994);\\
A. Datta, preprint UH-511-824-95 (1995).

\bibitem{hqet}
N. Isgur and M. Wise, Nucl. Phys. B348, 276 (1990);\\
H. Georgi,   Nucl. Phys. B348, 293 (1990);\\
T. Mannel, W. Roberts and Z. Ryzak,  Nucl. Phys. B355, 38 (1991);\\
T. Hussain, J.G. K\"orner, M. Kr\"amer and G. Thompson,
Z. Phys. C51, 321 (1991).

\bibitem{kk}
J. G. K\"orner and M. Kr\"amer, Phys. Lett. B275, 495 (1992).

\bibitem{sing}
R. Singleton, Phys. Rev.  D43, 2939 (1991).

\bibitem{helicity}
J. G. K\"orner and M. Kr\"amer, Z. Phys. C55, 659 (1992).

\bibitem{pdg}
Particle  Data  Group,  Phys. Rev. D50, 3$-I$ (1994).

\bibitem{burdman}
G. Burdman, E. Golowich, J.L. Hewett and S. Pakvasa,   
Phys. Rev. D52, 6383 (1995).

\bibitem{bvmd2}
N.G. Deshpande, X.-G. He and J. Trampetic, preprint OITS-564-REV,
Phys. Lett. B (to be published);\\
D. Atwood, B. Blok  and A. Soni, preprint SLAC-PUB-95-6635,
Inter. J. Mod. Phys. (to be published).

\bibitem{svmd}
G. Eilam, A. Ioannissian, R.R. Mendel and P. Singer, preprint TECHNION 95-18,
Phys. Rev.  D (to be published).

\bibitem{bvmd}
N.G. Deshpande, J. Trampetic and K. Panose, Phys. Lett. B214, 467 (1988);\\
E. Golowich and S. Pakvasa,  Phys. Rev.  {D51}, 1215 (1995);\\
H.-Y. Cheng, preprint IP-ASTP-23-94.

\bibitem{cheng3}
H.-Y. Cheng, C.-Y. Cheung, G.-L. Lin, Y.C. Lin, T.-M. Yan
and H.-L.  Yu, Phys. Rev.  D51, 1199 (1995).

\bibitem{vmd}
 K. Terasaki, Nuov. Cim.  {66A}, 475 (1981).

\bibitem{constitute}
T. Uppal and R.C. Verma, Phys. Rev.  D47, 2858 (1993).

\end{thebibliography}
\end{document}